\documentclass{iau}
\usepackage{graphicx}
\usepackage{floatrow}

\title[CMEs on Young Stars] 
{Coronal Mass Ejections and Angular Momentum Loss in Young Stars}

\author[Alicia Aarnio, Keivan Stassun, and Sean Matt]   
{Alicia N. Aarnio$^1$
\and Keivan G. Stassun $^{2,3}$
 \and Sean P. Matt$^4$}

\affiliation{$^1$Dept. of Astronomy, University of Michigan, 
Ann Arbor, MI, 48109, USA \\ email: {\tt aarnio@umich.edu}\\[\affilskip]
$^2$Dept. of Physics \& Astronomy, Vanderbilt University, 
Nashville, TN, 37235, USA \\[\affilskip]
$^3$Dept. of Physics, Fisk University, Nashville, TN, 37208 USA \\[\affilskip]
$^4$School of Physics, University of Exeter, \\  Stocker Road,
Exeter, EX4 4QL, UK }

\pubyear{2013}
\volume{300}  
\pagerange{}
\jname{Nature of prominences and their role in space weather}
\editors{A.C. Editor, B.D. Editor \& C.E. Editor, eds.}
\begin{document}

\maketitle

\begin{abstract}
In our own solar system, the necessity of understanding space weather is readily evident. Fortunately 
for Earth, our nearest stellar neighbor is relatively quiet, exhibiting activity levels several orders 
of magnitude lower than young, solar-type stars. In protoplanetary systems, stellar magnetic phenomena 
observed are analogous to the solar case, but dramatically enhanced on all physical scales: bigger, 
more energetic, more frequent. While coronal mass ejections (CMEs) could play a significant role in 
the evolution of protoplanets, they could also affect the evolution of the central star itself. To 
assess the consequences of prominence eruption/CMEs, we have invoked the solar-stellar connection to 
estimate, for young, solar-type stars, how frequently stellar CMEs may occur and their attendant mass 
and angular momentum loss rates. We will demonstrate the necessary conditions under which CMEs could 
slow stellar rotation. 
\end{abstract}

\firstsection 
\section{Introduction}
On young stars, we observe flares hundreds to ten thousand times more energetic and frequent than solar 
flares.
Along with energy scales greater by orders of magnitude, we also observe 
physical scales far greater than in the solar case: while solar prominences soar around 1 R$_{\odot}$ above 
the solar surface and CMEs launch from similar radii in the Sun's atmosphere, magnetic structures on 
T Tauri Stars (TTS, young solar analogs)--post-flare loops and prominences--can extend tens of stellar 
radii from the star's surface. 
The discovery of such large magnetic structures arose from solar-stellar analogy, applying solar flare 
models to the X-ray light curve data from young stars \cite[(e.g., Reale et al. 1998)]{Reale:1998}. 

In characterizing the solar-stellar connection, overwhelming evidence has been found in support of the idea 
that the fundamental physics of magnetic reconnection is the same, despite differences in stellar parameters 
(e.g., mass, radius, $B$, age). 
As such, we approach analysis of young stars' flares and CMEs under this supposition, and aim to assess how 
the physical properties of these events--and their frequency--may scale accordingly with stellar parameters. 
Ultimately, we seek to understand the consequences of exoplanetary space weather on protostellar systems and 
their forming planets.

\section{Estimating Angular Momentum Loss via Stellar CMEs}
The inference of magnetic loops many stellar radii in extent \cite[(Favata et al. 2005; hereafter F05)]{Favata:2005} inspired 
three questions: one, are these loops interacting with circumstellar disks? In \cite[Aarnio et al. (2010)]{Aarnio:2010}, 
we did not find evidence for this. Second, if there is not a star-disk link, how do the loops remain stable for the 
multiple rotation periods over which the X-ray flares are observed to decay? We showed in 
\cite[Aarnio et al. (2012)]{Aarnio:2012a} that when modeled as hot prominences, the addition of a scaled-up wind consistent with 
TTS observations provided sufficient support for the loops to be stable. Finally, here 
\cite[(and in Aarnio, Matt, \& Stassun, 2012; hereafter AA12)]{Aarnio:2012b} we address the third question of what happens 
when stability is lost: at many stellar radii, is the specific angular momentum shed significant enough to slow stellar rotation? 

In order to estimate the effects of eruptive prominences and stellar CMEs on the rotation of young stars, we must 
procure two ingredients: the mass lost via these events, and their frequency of occurrence. Despite ongoing and historical 
efforts to observe stellar CMEs, we lack definitive detections and thus frequency distributions. It is known that at 
times, magnetic reconnection on the Sun will produce both a flare and an associated CME; for stars, the 
flare is the observable quantity, and so we characterize stellar CME frequency by using stellar flare frequency as a proxy. 

\begin{figure}[b]
\floatbox[{\capbeside\thisfloatsetup{capbesideposition={right,top},capbesidewidth=2.3in}}]{figure}[\FBwidth]
{\caption{CME mass/flare energy relationship of AA11 shown with TTS post-flare loop masses (black diamonds) derived 
in AA12 plotted as a function of the flare's energy. Point size denotes the mass of the star on which the flare was 
observed. 
Black dotted lines show predicted post-flare stellar loop masses (Eqn. \ref{eqn1} and discussion in text) 
for a range of observed densities from 10$^{10}$-10$^{12}$ cm$^{-3}$ and assuming a confining field strength of 50G.
Gray, solid-lined boxes denote the range of observed X-ray flare energies 
\cite[(Maggio et al. 2000, Collier Cameron et al. 1988)]{Maggio:2000,CCameron:1988}
and cool 
H$\alpha$ prominence mass estimates \cite[(Collier Cameron \& Robinson, 1989a,b)]{CCameron:1989a,CCameron:1989b} 
for AB Dor and Speedy Mic, 20-50Myr K dwarfs.
}   \label{fig1}}
{\includegraphics[width=2.5in,angle=90,trim=0.6cm 0.8cm 1cm 1.8cm,clip=true]{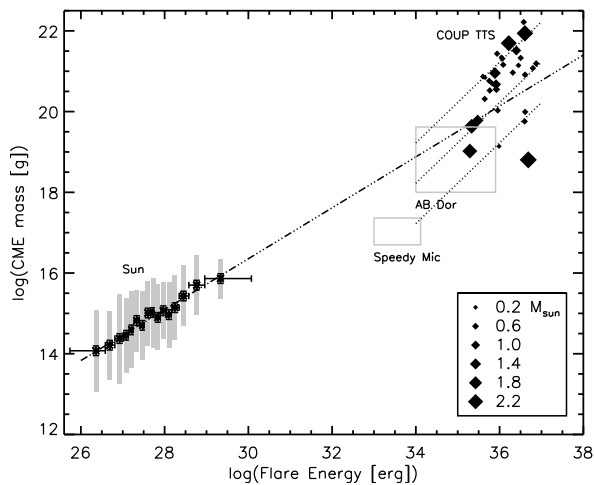}}
\end{figure}

In Fig. \ref{fig1}, we show our solar flare energy/CME mass relationship \cite[(Aarnio et al. 2011, hereafter AA11)]{Aarnio:2011} 
extrapolated up to the energies of young 
stellar flares. We calculate loop masses for the 32 ``superflaring'' stars from the \textit{Chandra} Orion Ultradeep 
Project \cite[(Getman et al. 2005)]{Getman:2005} from the parameters reported by F05. Interestingly, these loop masses 
are close in 
parameter space to the extrapolated solar relationship. This is perhaps unsurprising, as the plasma confined in a 
post-flare loop has properties which relate to the energy of the flare. In AA11, we 
found that for associated flares and CMEs, that is to say, flares and CMEs which likely originated from a shared 
magnetic reconnection event, the CME mass and flare energy were related. As such, the post-flare loop mass and mass 
of an associated CME should then also be related.

From \cite[Shibata \& Yokoyama (2002)]{Shibata:2002}, we can estimate the total energy released by a flare is related 
to the total magnetic energy in a flare loop:
\begin{equation}\label{eqn1}
E_{mag} = \frac{B^2 L^3}{8 \pi}.
\end{equation}
If we substitute the loop volume expressed in terms of mass (i.e., $L^3 \sim V=m_{\rm loop}/\rho$), we find a relationship 
between the total flare energy and the mass confined in the magnetic loop (dashed, parallel lines in Fig. \ref{fig1}). 
Here, to be consistent with the analysis of F05, we assume an Euclidian loop filling but do note that recent solar 
X-ray flare imaging has indicated that a fractal scaling of V(L)$\propto$L$^{2.4}$ is likely a more accurate 
characterization \cite[(e.g., Aschwanden, Stern, \& G\"{u}del, 2008)]{Aschwanden:2008}.

It is remarkable that the post-flare loop masses even lie near the extrapolated solar CME mass/flare energy relationship, 
several orders of magnitude away in parameter space. 
In Fig. \ref{fig1}, we have also shown representative ranges of X-ray flare energy and prominence mass for two K dwarfs 
intermediate in age to the TTS sample and the Sun; with an eruptive prominence thought to be the core of a CME, these mass 
ranges likely represent lower limits on the range of CME masses on these stars. 
In the following calculations, to represent a fiducial TTS case, we will simply extrapolate the solar relationship to 
generate a stellar CME mass distribution.

\begin{figure}
\floatbox[{\capbeside\thisfloatsetup{capbesideposition={left,top},capbesidewidth=2in}}]{figure}[\FBwidth]
{\caption{Flare frequencies for the Sun, M dwarfs \cite[(Hilton et al. 2011)]{Hilton:2011}, 
TTS in the ONC \cite[(Albacete Colombo et al. 2007)]{Colombo:2007}, and active, main sequence G stars 
from Kepler \cite[(Maehara et al. 2012)]{Maehara:2012}. Interestingly, the active G stars are almost indistinguishable 
from the TTS. Note the solar and TTS frequencies are derived from X-ray flare data, while the M dwarf and G stars are 
optical flare frequencies.}
   \label{fig2}}
{\includegraphics[width=2.0in,angle=90,clip=true,trim=0.5cm 0.5cm 0.5cm 1cm]{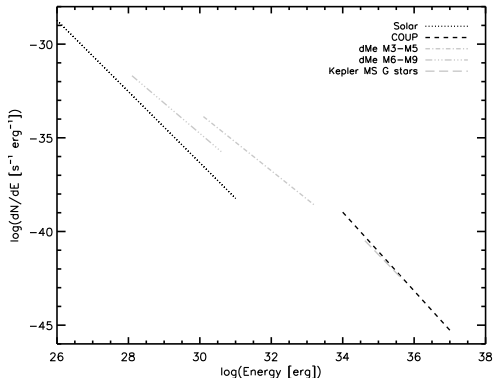} }
\end{figure}

In Fig. \ref{fig2}, 
we show the frequency distributions for flares observed on the Sun, TTS, M dwarfs, and active, main sequence G stars. For this 
work, we use the TTS frequency distribution. Clearly, not all 
CMEs are flare-associated, nor are all flares CME-associated; AA11 found, however, that the 
association fraction increases with increasing flare energy, so for young stars for which we observe flares 
several orders of magnitude more energetic than in the solar case, we simply assume this association fraction to be of order 
unity. 

\subsection{Angular momentum loss}

In AA12, we extrapolate the solar CME mass/flare energy relationship 
(Fig. \ref{fig1}) to TTS flare energies and frequencies (Fig. \ref{fig2}) and construct a CME frequency distribution 
as a function of CME mass. Given the observational completeness limits on the flare distributions that went into the 
CME distribution, we derive lower and upper limits on the mass loss rate by empirical (integrating the distribution) 
and analytical (integrating a fit to the distribution) means. 
The range of mass loss rates we estimate for the TTS 
case is 10$^{-12}$-10$^{-9}$ M$_{\odot}$ yr$^{-1}$.

To assess the torque applied against stellar rotation by these CMEs, we apply stellar wind models with mass loss rates 
set as determined above. Given the episodic nature of CMEs, we included an efficiency parameter to account for the fact that 
steady-state winds are more efficient at removing angular momentum than ``clumpy'' winds (cf. AA12 and references therein). 
We adopt a dipolar field with strength 600G, consistent with observations of 
TTS fields, and allow the stellar radius to contract as stellar evolution models predict.

In a protostellar system, multiple torques act simultaneously to spin up and spin down the star. In this analysis, we 
compared spin up due to contraction and spin down due to mass loss from stellar CMEs to see if, at any point in the 
pre-main sequence our fiducial TTS could have its rotation slowed due to CMEs. Comparing parameters of efficiency and 
the range of mass loss rates we calculated, it became clear that only towards the end of the pre-main sequence phase 
(ages $\gtrsim$6 Myr)
could a very efficient, high CME mass loss rate begin to counteract spin up from contraction. We have left out factors 
such as spin up from accretion and mass loss via stellar wind; \cite[Matt \& Pudritz (2005)]{Matt:2005} explore these 
two torques in depth and the necessary conditions for an accretion powered stellar wind to slow stellar rotation.

\section{Discussion}\label{disc}

We have shown that for young, solar-type stars, spin down due to CMEs might play a significant role in stellar rotation 
evolution after the star has ceased accreting. Our Figs. \ref{fig1} and \ref{fig2} illustrate a critical selection effect 
in performing this kind of calculation: we only have data for 
the most active young stars, or the star conveniently located at 1 AU. There is a dearth of data for older, less active stars, 
and we suggest that filling in the gaps in flare X-ray energy could trace age evolution in these parameter spaces. 
The addition of data from the $\sim$20-50 Myr old K dwarfs in Fig. \ref{fig1} hints at this, but more data are needed to conclusively 
show age dependence. In both figures, we have taken care to specify the masses of the stars involved: how would evolution with stellar 
age look in these parameter spaces as a function of stellar mass? While the fundamental physics are the same, the scaling could 
change, and the ramifications certainly would. For low-mass stars in particular, high activity levels are observed for longer 
fractions of the stars' lives; this could have grave implications for exoplanets as these stars' habitable zones could be 
within range of extreme exo-weather.

\acknowledgments
A.~N.~A. thanks K. Shibata for helpful discussion regarding stellar post-flare loops.



%
%

\end{document}